\documentclass[12pt,fleqn]{article}
\usepackage{amsmath,amssymb,graphicx,epsfig}

\def\bea{\begin{eqnarray}}
\def\eea{\end{eqnarray}}
\def\be{\begin{equation}}
\def\ee{\end{equation}}
\def\ba{\begin{array}}
\def\ea{\end{array}}
\def\nn{\nonumber}
\def\a{& \hspace{-10pt}}
\def\b{& \hspace{-7pt}}

\def\s#1{\text{\small $#1$}}
\def\t#1{\text{\tiny $#1$}}

\newcommand{\bal}{\begin{aligned}}
\newcommand{\eal}{\end{aligned}}
\newcommand{\e}{\text{eff}}

\font\tenrsfs=rsfs10
\font\sevenrsfs=rsfs7
\font\fiversfs=rsfs5
\newfam\rsfsfam
\textfont\rsfsfam=\tenrsfs
\scriptfont\rsfsfam=\sevenrsfs
\scriptscriptfont\rsfsfam=\fiversfs
\def\mathscr#1{{\fam\rsfsfam\relax#1}}

\begin{document}

\thispagestyle{empty}

\begin{center}

$\;$

\vspace{1cm}

{\LARGE \bf Effects of heavy modes on vacuum \\[2mm]
stability in supersymmetric theories}

\vspace{1.3cm}

{\large {\bf Leonardo~Brizi} and {\bf Claudio~A.~Scrucca}}\\[2mm] 

\vspace{0.4cm}

{\large \em Institut de Th\'eorie des Ph\'enom\`enes Physiques\\ 
Ecole Polytechnique F\'ed\'erale de Lausanne\\ 
CH-1015 Lausanne, Switzerland\\}

\vspace{0.2cm}

\end{center}

\vspace{1cm}

\centerline{\bf \large Abstract}
\begin{quote}

We study the effects induced by heavy fields on the masses of light 
fields in supersymmetric theories, under the assumption that the heavy 
mass scale is much higher than the supersymmetry breaking scale. 
We show that the square-masses of light scalar fields can get two 
different types of significant corrections when a heavy multiplet is 
integrated out. The first is an indirect level-repulsion effect, which 
may arise from heavy chiral multiplets and is always negative. 
The second is a direct coupling contribution, which may 
arise from heavy vector multiplets and can have any sign. 
We then apply these results to the sGoldstino mass and study 
the implications for the vacuum metastability condition. 
We find that the correction from heavy chiral multiplets is always 
negative and tends to compromise vacuum metastability, whereas the 
contribution from heavy vector multiplets is always positive and tends 
on the contrary to reinforce it. These two effects are controlled respectively
by Yukawa couplings and gauge charges, which mix one heavy and two 
light fields respectively in the superpotential and the K\"ahler potential.
Finally we also comment on similar effects induced in soft scalar
masses when the heavy multiplets couple both to the visible and the 
hidden sector.

\vspace{5pt}
\end{quote}

\renewcommand{\theequation}{\thesection.\arabic{equation}}

\newpage

\setcounter{page}{1}

\section{Introduction}
\setcounter{equation}{0}

In supersymmetric theories, vacua that preserve supersymmetry are 
automatically stable, whereas vacua that break supersymmetry are
not guaranteed to be stable. In order to assess stability, one then has 
to study the mass matrix of scalar fluctuations around the vacuum and 
check that it is positive definite. It was however shown in \cite{GRS1,GRS2,GRS3} 
(see also \cite{DD} for a related analysis), by looking at the sGoldstino 
direction, that there exists a simple necessary condition for metastability 
depending on the sectional curvature of the scalar manifold along the 
supersymmetry breaking direction. Moreover, it has been further argued 
in \cite{CGGLPS,CGGPS} that this condition becomes also sufficient if 
for a given K\"ahler potential $K$ one allows the superpotential $W$ to 
be adjusted. These results are quite helpful for discriminating between 
theories where metastable vacua may exist and theories where they 
cannot exist, by looking only at $K$ and not at $W$. 
A comprehensive review of these results and some extensions of them 
within rigid supersymmetry can be found in \cite{JS}. 

In some cases, like for instance for the moduli sector of string models 
where supersymmetry is supposed to be spontaneously broken, one 
may be interested in studying the possibility that some of the fields 
are stabilized in a supersymmetry-breaking way with a small mass, 
whereas the remaining fields are stabilized in a supersymmetry-preserving 
way with a large mass. One may then study the low-energy dynamics and 
in particular the question of vacuum metastability within a supersymmetric 
effective theory obtained by integrating out the heavy multiplets. The way in 
which this can be done in a manifestly supersymmetric way is well known,
see for instance \cite{IS,ADM}, and turns out to hold true also in the presence
of gravity \cite{BGRS}.\footnote{See also the earlier work \cite{dA} where 
this question was raised and the works \cite{C1,C2} studying it in the 
case of effective theories describing string models with fluxes.}
At leading order in the low-energy expansion in the number of derivatives,
fermions and auxiliary fields, the basic recipe is that chiral and vector superfields 
can be integrated out by using an approximate equation of motion corresponding 
to imposing stationarity of $W$ and $K$ respectively. One may then ask the 
practical question of what is the effect of heavy modes on the light masses, 
and in particular whether the induced corrections tend to improve or to worsen 
the situation concerning metastability of the vacuum. More specifically, it would 
be very valuable to have some criterium to distinguish situations where 
the effect of heavy modes on the scalar square-masses are negative, 
and must therefore necessarily be computed to be able to assess vacuum 
stability, from situations where this effect is positive and can thus be safely 
ignored to check vacuum metastability. To derive such a criterium, we shall 
study in some detail the structure and the sign of the effect induced by 
heavy modes on the sGoldstino mass, which captures the crucial condition 
for achieving metastability. For simplicity we shall restrict to rigid supersymmetry,
but the extension of supergravity is straightforward, since as explained 
in \cite{BGRS} the two steps of integrating out heavy multiplets and adding the 
coupling to gravity commute at leading order in the low-energy expansion.

In order to illustrate the basic point that we want to make, let us consider 
a generic theory involving both light and heavy modes $l^i$ and $h^\alpha$ 
that interact among each other. For simplicity, we shall think of these as 
real scalar fields in a non-supersymmetric theory, but the results are clearly 
more general. In such a situation, one may define a low-energy effective 
theory for the light modes $l^i$ by integrating out the heavy modes $h^\alpha$.
At lowest order in the low-energy expansion, this can be done by requiring 
stationarity of the potential energy $V$ with respect to the heavy modes and 
solving the equation $V_\alpha = 0$. This determines $h^\alpha = h^\alpha(l)$. 
By differentiating the stationarity equation with respect to the light fields, one also 
deduces that $\partial_i h^\alpha = - V^{\alpha \beta}_{\rm inv} V_{\beta i}$,
where $V^{\alpha \beta}_{\rm inv}$ denotes the inverse of $V_{\alpha \beta}$
as a matrix. The effective Lagrangian for the low-energy theory is then obtained by 
substituting back this solution into the original Lagrangian. For the wave-function
factor and the potential, one easily obtains
$g_{ij}^{\rm eff}(l) = (g_{ij} + \partial_i h^\alpha g_{\alpha j} 
+ \partial_j h^\beta g_{i \beta} + \partial_i h^\alpha \partial_j h^\beta g_{\alpha \beta})(l,h(l))$
and $V^{\rm eff} (l) = V(l,h(l))$. The light masses may finally be derived by 
computing derivatives of $V^{\rm eff}$. Using the chain rule, these can be 
related to derivatives of $V$. One finds $V^{\rm eff}_i = V_i$ 
and $V^{\rm eff}_{ij} = V_{ij} - V_{i \alpha} V^{\alpha \beta}_{\rm inv} V_{\beta j}$, 
so that the light masses $m^{2{\rm eff}}_{ij} = V^{\rm eff}_{ij}$ are given by the 
following expression in terms of the light, heavy and mixing blocks 
$m^2_{ij} = V_{ij}$, $M^2_{\alpha \beta} = V_{\alpha \beta}$ and 
$\mu^2_{i \alpha} = V_{i \alpha}$ of the full mass matrix:
\be
m^{2{\rm eff}}_{ij} = m^2_{ij} - \mu^2_{i \alpha} M^{-2\alpha \beta} \! \mu^2_{\beta j} \,.
\label{ml}
\ee
This expressions is easily seen to coincide with the mass matrix of light states 
obtained by diagonalizing the full mass matrix of the microscopic theory 
at leading order in an expansion in powers of the inverse heavy mass matrix. 
The formula (\ref{ml}) moreover shows that integrating out the heavy modes 
generically gives two types of effects on the masses of the light modes. The first 
is a direct effect hidden in the first term on the right hand side and is due to the 
fact that the light block of the mass matrix $m^2_{ij}$ gets influenced by the 
coupling to the heavy modes. It has a sign that depends on the form of the 
couplings between light and heavy modes. The second is an indirect effect 
described by the second term on the right-hand side and is due to the fact 
that the presence of an off-diagonal block in the mass matrix mixing light and 
heavy fields makes the true light mass matrix differ from the original light block.
It has a sign that is manifestly always negative. In parallel with what happens to 
a quantum mechanical system with two separated sets of low and high energy 
levels, we see that there is a direct effect correcting significantly the light energy 
levels and negligibly the heavy ones, which is due to diagonal interactions and 
can have any sign, and an indirect level-repulsion effect that further splits apart 
the two sets of levels, which is due to off-diagonal interactions and has a definite sign. 

In this work, we shall consider $N=1$ supersymmetric theories and compute the 
detailed form of the effective mass matrix for light scalar fields belonging to chiral 
multiplets in the two cases where the heavy modes that are integrated out are 
respectively chiral and vector multiplets. More precisely, we shall focus on the 
mass along the sGoldstino direction, to extract the metastability condition. It turns out 
that two radically different results occur in these two situations. In the case of heavy 
chiral multiplets only the indirect level-repulsion effect generically arises with a 
non-negligible size. The correction is always negative and thus dangerous, as 
suggested from the arguments in \cite{L}. We will derive its general form and show that 
it is controlled by the mixed third derivatives of $W$. In the case of heavy vector 
multiplets, on the other hand, only the direct effect occurs. Moreover the correction 
turns out to be always positive and therefore harmless, as already argued 
in \cite{GRS3}. We will rederive more precisely its form, which is controlled by 
the mixed third derivatives of $K$. 

\section{Models with chiral multiplets}
\setcounter{equation}{0}

Let us start by considering the simplest case of $N=1$ theories with only
chiral multiplets $\Phi^I$. The most general two-derivative Lagrangian is specified 
in terms of a real K\"ahler potential $K$ and a holomorphic superpotential $W$, and reads:
\bea
\mathcal{L} = \int \! d^4 \theta\, K(\Phi,\bar \Phi) + \int \! d^2 \theta\, W(\Phi) + {\rm h.c.} \,.
\eea
In components, this gives $\mathcal{L} = T - V$ where
\bea
T \b=\b-g_{I \bar J} \, \partial_\mu \phi^I \partial^\mu \bar \phi^{\bar J}
-ig_{I \bar J} \, \psi^I \big(\partial\!\!\!/\bar \psi^{\bar J} + 
\Gamma^{\bar J}_{\bar M \bar N} \, \partial\!\!\!/ \bar \phi^{\bar M} \bar \psi^{\bar N} \big) \,, \\
V \b=\b g^{I \bar J} \, W_I \bar W_{\bar J} 
+ \s{\frac 12} \nabla_I W_J \, \psi^I \psi^J + {\rm h.c.}
- \s{\frac 14} R_{I \bar J K \bar L} \, \psi^I \psi^K \bar \psi^{\bar J} \bar \psi^{\bar L} \,,
\eea

A vacuum is defined by constant values of the scalars $\phi^I$ and 
vanishing values of the fermions $\psi^I$, such that $V$ is stationary. Supersymmetry is 
spontaneously broken whenever some of the auxiliary fields $F^I$ have non-vanishing 
values. The form of these auxiliary fields is given by
\be
F^I =  - g^{I \bar J} \, \bar W_{\bar J} \,.
\ee
The stationarity condition implies moreover that 
\be\label{stat}
\nabla_I W_J \,F^J = 0 \,.
\ee
The masses for the scalar and fermion fields describing fluctuations around the vacuum 
are then found to be given by
\bea\label{mo}
\a\a m^2_{0 I \bar J} = \nabla_I W_K \nabla_{\bar J} \bar W^K - R_{I \bar J K \bar L} \, F^K \bar F^{\bar L} \,, \\
\a\a m^2_{0 I J} = - \nabla_I \nabla_J W_K \, F^K \,,
\eea
and
\bea
\a\a m_{1/2 IJ} = \nabla_I W_J \,.
\eea
We see from the above expressions that the supersymmetric part of the mass is 
controlled by the quadratic terms in the superpotential and given by $W_{IJ}$.

The direction $F^I$ in field space is special. For fermions it defines the Goldstino 
$\eta = \bar F_I \psi^I$, which is massless and represents the Goldstone
mode of broken supersymmetry: $m_\eta = 0$. 
For scalars it defines instead the sGoldstino $\varphi = \bar F_I \phi^I$, which 
describes two real scalar fields with masses that are entirely controlled by 
supersymmetry breaking effects. One may then look at the average of these
two masses, which is defined as
\be
m_\varphi^{2} = \frac 
{m_{0I \bar J}^{2} F^{I} \bar F^{\bar J}}
{F^K \bar F_K} \,.
\label{mmic}
\ee
A simple computation shows that this is given by \cite{GRS1,JS}
\be\label{msgm}
m_{\varphi}^2 = R \,F^I \bar F_I \,,
\ee
where
\be
R = - \frac {R_{I \bar J K \bar L} F^I \bar F^{\bar J} F^K \bar F^{\bar L}}{(F^M \bar F_M)^2} \,.
\label{RF}
\ee

From this result it follows that a necessary condition for not having a tachyonic 
mode is that the holomorphic sectional curvature $R$ be positive \cite{GRS1,GRS2}. 
This necessary condition becomes also sufficient if for a given $K$ 
one allows $W$ to be adjusted \cite{CGGLPS}. Indeed, 
at the stationary point one may tune $W_I$ to maximize the average sGoldstino mass, 
$W_{IJ}$ to make the other masses arbitrarily large, and $W_{IJK}$ to set the splitting 
between the two sGoldstino masses to zero. Moreover, in such a situation one can prove
that the two real sGoldstino modes become degenerate mass eigenstates \cite{CGGPS}. 

\subsection{Integrating out heavy chiral multiplets}

Let us now consider a situation where the chiral multiplets $\Phi^I$ split into 
a set of light multiplets $\Phi^i$ parametrizing the low-energy theory and 
a set of heavy multiplets $\Phi^\alpha$ with a large supersymmetric mass 
$W_{\alpha\beta}$ to be integrated out. In order to distinguish light from 
heavy multiplets in a sensible way, we must assume that the supersymmetric 
mass mixing $W_{i \alpha}$ between them is not too large. 
In the following, we shall denote these heavy and mixing blocks of the 
supersymmetric mass matrix in the following way:
\be
M_{\alpha \beta} = W_{\alpha\beta} \,,\;\; 
\mu_{i\alpha} = W_{i \alpha} \,.
\ee
The most relevant interactions for our purposes will be the cubic terms in $W$, 
namely the Yukawa couplings
\be
\lambda_{\alpha i j} = W_{\alpha i j} \,,\;\; 
\lambda_{\alpha \beta j} = W_{\alpha \beta j} \,,\;\; 
\lambda_{\alpha \beta \gamma} = W_{\alpha \beta \gamma} \,.
\ee

At leading order in the low-energy expansion in number of derivatives, fermions and 
auxiliary fields, the low-energy effective theory can be obtained in component fields 
by imposing stationarity of $V$ with respect to each heavy field and substituting back 
the solution into the original Lagrangian. Equivalently, this effective theory can be derived 
directly in superfields, by demanding the stationarity of $W$ with respect to each heavy 
chiral multiplet. For convenience, we shall assume without loss of generality normal 
coordinates in the microscopic theory around the point under consideration. This substantially 
simplifies the computations, although the effective theory does not automatically inherit 
normal coordinates, due to the corrections induced to the K\"ahler metric.

The corrections due to the supersymmetric mass mixing between heavy and light multiplets 
are encoded in the following small dimensionless matrix:
\be
\epsilon^\alpha_i = - M^{-1 \alpha \beta} \! \mu_{\beta i} \,.
\ee
It should be emphasized that it is always possible to perform a holomorphic field redefinition 
in such a way to diagonalize the supersymmetric mass matrix $W_{IJ}$ at a given point in field space,
thereby setting $\epsilon^\alpha_i$ to zero. This means that all the effects depending on 
$\epsilon^\alpha_i$ only serve to compensate a choice of light and heavy fields that does not 
exactly diagonalize the supersymmetric part of the mass matrix, and therefore do not represent
genuine non-trivial corrections. Moreover, since $\epsilon^\alpha_i$ must be small, these effects 
are anyhow quantitatively irrelevant. We may then set $\epsilon^\alpha_i = 0$ by suitably choosing 
the fields. We shall however keep $\epsilon^\alpha_i \neq 0$ during the computations to verify 
more explicitly the above claims and set $\epsilon^\alpha_i = 0$ only at the very end. We can 
anticipate that all the tensorial quantities characterizing the light fields will receive additional 
contributions coming from heavy indices converted to light indices through the matrix $\epsilon^\alpha_i$. 
This leads us to introduce already at this stage the following deformed tensors:
\bea
g_{i \bar \jmath}^\epsilon \b=\b g_{i \bar \jmath} + \epsilon_i^\alpha g_{\alpha \bar \jmath} 
+ \bar \epsilon_{\bar \jmath}^{\bar \beta} g_{i \bar \beta} 
+ \epsilon_i^\alpha \bar \epsilon_{\bar \jmath}^{\bar \beta} g_{\alpha \bar \beta} \,, \\
\lambda^\epsilon_{\alpha i j} \b=\b \lambda_{\alpha i j} + \epsilon_i^\beta \lambda_{\alpha \beta j} 
+ \epsilon_j^\gamma \lambda_{\alpha i \gamma} + \epsilon_i^\beta \epsilon_j^\gamma \lambda_{\alpha \beta \gamma} \,, \\
R^\epsilon_{i \bar \jmath k \bar l} \b=\b R_{i \bar \jmath k \bar l} 
+ \epsilon_i^\alpha R_{\alpha \bar \jmath k \bar l} 
+ \bar \epsilon_{\bar \jmath}^{\bar \beta} R_{i \bar \beta k \bar l} 
+ \epsilon_k^\gamma R_{i \bar \jmath \gamma \bar l}
+ \bar \epsilon_{\bar l}^{\bar \delta} R_{i \bar \jmath k \bar \delta} 
+ \epsilon_i^\alpha \bar \epsilon_{\bar \jmath}^{\bar \beta} R_{\alpha \bar \beta k \bar l}  \nn \\
\b\;\b +\, \epsilon_i^\alpha  \epsilon_k^\gamma R_{\alpha \bar \jmath \gamma \bar l}  
+ \epsilon_i^\alpha \bar \epsilon_{\bar l}^{\bar \delta} R_{\alpha \bar \jmath k \bar \delta}
+ \bar \epsilon_{\bar \jmath}^{\bar \beta} \epsilon_k^\gamma R_{i \bar \beta \gamma \bar l}  
+ \bar \epsilon_{\bar \jmath}^{\bar \beta} \bar \epsilon_{\bar l}^{\bar \delta} R_{i \bar \beta k \bar \delta}  
+ \epsilon_k^\gamma \bar \epsilon_{\bar l}^{\bar \delta} R_{i \bar \jmath \gamma \bar \delta} \nn \\
\b\;\b +\, \bar \epsilon_{\bar \jmath}^{\bar \beta} \epsilon_k^\gamma \bar \epsilon_{\bar l}^{\bar \delta} R_{i \bar \beta \gamma \bar \delta} 
+ \epsilon_i^\alpha \epsilon_k^\gamma \bar \epsilon_{\bar l}^{\bar \delta} R_{\alpha \bar \jmath \gamma \bar \delta} 
+ \epsilon_i^\alpha \bar \epsilon_{\bar \jmath}^{\bar \beta} \bar \epsilon_{\bar l}^{\bar \delta} R_{\alpha \bar \beta k \bar \delta} 
+ \epsilon_i^\alpha \bar \epsilon_{\bar \jmath}^{\bar \beta} \epsilon_k^\gamma R_{\alpha \bar \beta \gamma \bar l} \nn \\
\b\;\b +\, \epsilon_i^\alpha \bar \epsilon_{\bar \jmath}^{\bar \beta} \epsilon_k^\gamma \bar \epsilon_{\bar l}^{\bar \delta} 
R_{\alpha \bar \beta \gamma \bar \delta} \,.
\eea
Finally, we shall define the following quantity for later use, which characterizes the heavy block 
$W_{\alpha I} g^{I \bar J} \bar W_{\bar J \bar \beta}$ of the square of the supersymmetric mass matrix:
\be
|M^\epsilon|^2_{\alpha \bar \beta} = M_{\alpha \gamma} 
\big(g^{\gamma \bar \delta} + \epsilon_i^\gamma g^{i \bar \delta} 
+ \bar \epsilon_{\bar \jmath}^{\bar \delta} g^{\gamma \bar \jmath} 
+ \epsilon_i^\gamma \bar \epsilon_{\bar \jmath}^{\bar \delta} g^{i \bar \jmath} 
\big) \bar M_{\bar \delta \bar \beta} \,.
\ee

In the following, we shall compute within the component approach the average sGoldstino
mass in the low-energy effective theory, defined at a stationary point as
\be
m_\varphi^{2{\rm eff}} = \frac 
{m_{0i \bar \jmath}^{2{\rm eff}} F^{i{\rm eff}} \bar F^{\bar \jmath{\rm eff}}}
{F^{k{\rm eff}} \bar F_k^{\rm eff}} \,.
\label{meff}
\ee
We shall then reproduce the same result within the superfield approach by first computing 
the Riemann tensor $R^\e_{i \bar \jmath k \bar l}$ of the effective theory at a generic point 
and then applying the standard expression for the sGoldstino mass at a stationary point 
within the effective theory, namely
\be
m_\varphi^{2{\rm eff}} = R^{\rm eff} F^{i{\rm eff}} \bar F_i^{\rm eff} \,, \label{meff1}
\ee
in terms of an effective sectional curvature
\be
R^{\rm eff} = - \frac {R_{i \bar \jmath k \bar l}^{\rm eff} F^{i{\rm eff}} \bar F^{\bar \jmath{\rm eff}} F^{k{\rm eff}} \bar F^{\bar l{\rm eff}}}
{(F^{m{\rm eff}} \bar F_m^{\rm eff})^2} \,. \label{meff2}
\ee

\subsection{Component approach}

Consider first the component approach. For simplicity we shall focus on the bosonic fields and 
discard fermions, since we are interested in computing effective scalar masses. At leading order
in the low-energy expansion, the values of the heavy scalar fields are defined by
\bea
\phi^\alpha \b=\b \phi^\alpha (\phi^i,\bar \phi^{\bar \imath}) \;\;\text{solution of}\;\; 
V_\alpha (\phi^i,\bar \phi^{\bar \imath}, \phi^\alpha, \bar \phi^{\bar \alpha}) = 0 \,. \label{aphi}
\eea
At leading order in the number of auxiliary fields, this stationarity condition
implies that $W_{\alpha I} \bar W^I = 0$ and gives the following values for the 
heavy auxiliary fields:
\bea
F^\alpha \b=\b \epsilon^\alpha_i F^i \label{aF} \,.
\eea
The effective theory for the light fields is then obtained by substituting these 
expressions for $\phi^\alpha$ and $F^\alpha$ into the original Lagrangian.

To derive the effective theory, we will need to compute the derivatives of the heavy fields 
$\phi^\alpha$ and $\bar \phi^{\bar \alpha}$ with respect to the light fields $\phi^i$. These 
can be deduced by differentiating the stationarity conditions with respect to the light fields.
One finds:
\bea
\a\a \frac {\partial \phi^\alpha}{\partial \phi^i} = - M_0^{-2 \alpha \bar \beta} \mu^2_{0 \bar \beta i} 
- M_0^{-2 \alpha \beta} \mu^2_{0 \beta i} \,, \label{der1} \\
\a\a \frac {\partial \bar \phi^{\bar \alpha}}{\partial\phi^i} = - M_0^{-2 \bar \alpha \beta} \mu^2_{0 \beta i} 
- M_0^{-2 \bar \alpha \bar \beta} \mu^2_{0 \bar \beta i} \,. \label{der2}
\eea
Here $M_0^2$ and $\mu_0^2$ represent the heavy and off-diagonal blocks of the complete scalar mass 
matrix of the microscopic theory. Notice that $\mu_0$ and $M_0$ differ from $\mu$ and $M$, since the 
former refer to the full mass matrix whereas the latter parametrize only its supersymmetric
part. At quadratic order in the auxiliary fields one finds:
\bea
M_0^{-2\alpha\bar\beta} \b=\b V_{\rm inv}^{\alpha\bar\beta}
+ V_{\rm inv}^{\alpha\bar\gamma} V_{\bar \gamma \bar \delta} V_{\rm inv}^{\bar\delta\sigma}
V_{\sigma \tau} V_{\rm inv}^{\tau\bar\beta} \,, \label{f1} \\
M_0^{-2 \alpha\beta} \b=\b -V_{\rm inv}^{\alpha\bar\gamma} V_{\bar \gamma \bar \delta}
V_{\rm inv}^{\bar\delta\beta} \,. \label{f2}
\eea

The effective K\"ahler metric of the light fields can be determined by looking at the scalar 
kinetic terms and substituting the values of the heavy scalar fields. One may in this case 
work at leading order in the auxiliary fields, since these terms already involve two derivatives. 
Focusing also on the leading order in the light masses and the heavy-light mass mixing, the 
relations (\ref{der1}) and (\ref{der2}) then simplify to $\partial_i \phi^\alpha = \epsilon^\alpha_i$ 
and $\partial_i \bar \phi^{\bar \alpha} = 0$. Using these expressions, which actually turn out to 
be correct even at order $\epsilon^2$, one finds that the kinetic term can be rewritten in the 
standard supersymmetric form with an effective K\"ahler metric given by
\bea
g^ {\rm eff}_{i\bar\jmath} \b=\b g_{i \bar \jmath}^\epsilon \label{keff} \,.
\eea

The effective mass matrix of the light scalar fields can on the other hand be determined 
by using the supersymmetric generalization of the expression (\ref{ml}), which can 
be derived by using the same logic. As in the general non-supersymmetric case, the 
result corresponds to a perturbative diagonalization of the full scalar mass matrix, 
at leading order in the inverse mass matrix of the heavy scalars. Denoting with $m_0^2$
the light block of the scalar mass matrix, one finds:
\bea
m^{2{\rm eff}}_{0 i\bar\jmath} \b=\b m^2_{0 i\bar\jmath}
- \mu^2_{0i\bar\alpha} M_0^{-2\bar \alpha \beta} \! \mu^2_{0\beta\bar\jmath}
- \mu^2_{0i\bar\alpha} M_0^{-2 \bar \alpha \bar \beta} \! \mu^2_{0\bar\beta\bar\jmath} \nn \\
\b\;\b -\, \mu^2_{0i\alpha} M_0^{-2 \alpha \beta} \! \mu^2_{0\beta\bar\jmath}
- \mu^2_{0i\alpha} M^{-2 \alpha \bar \beta} \! \mu^2_{0\bar\beta\bar\jmath} 
\label{mibj} \,, \\
m^{2{\rm eff}}_{0 i j} \b=\b m^2_{0 i j}
- \mu^2_{0i\bar\alpha} M_0^{-2 \bar \alpha \bar \beta} \! \mu^2_{0\bar \beta j}
- \mu^2_{0i\bar\alpha} M_0^{-2 \bar \alpha \beta}\! \mu^2_{0\beta j} \nn \\
\b\;\b -\, \mu^2_{0i\alpha} M_0^{-2 \alpha \bar \beta}\! \mu^2_{0\bar \beta j}
- \mu^2_{0i\alpha} M_0^{-2 \alpha\beta}\! \mu^2_{0\beta j}
\label{mij} \,.
\eea
Let us now focus on the Hermitian block $m^{2 \e}_{0 i \bar \jmath}$. 
Using eqs.~(\ref{f1}) and (\ref{f2}) in the formula (\ref{mibj}), and restricting to 
terms that are at most quadratic in the auxiliary fields as demanded by supersymmetry 
at the two-derivative level, we see that there are three kinds of effects coming 
from the four correction terms. 
The first type involves second derivatives of $W$ and no auxiliary fields,
and comes only from the first correction term. 
The second type involves the Riemann tensor and two auxiliary fields,
and comes again only from the first correction term. 
The third type involves third derivatives of $W$ and two auxiliary fields,
and comes from all four correction terms. All together, these three effects
give a negative level-repulsion correction with respect to $m^2_{0 i \bar \jmath}$.

Let us now compute more specifically the average sGoldstino mass $m_\varphi^{2{\rm eff}}$ 
defined by eq.~(\ref{meff}) at a stationary point of the effective theory and compare it to 
its analogue $m_\varphi^2$ defined by eq.~(\ref{mmic}) in the microscopic theory.
Recall that we are using normal coordinates, so that $g_{i \bar \jmath} = \delta_{i \bar \jmath}$ 
and $g_{i \bar \jmath}^{\rm eff} = \delta_{i \bar \jmath} + \epsilon_i^\alpha \bar \epsilon_{\bar \jmath}^{\bar \alpha}$.
The first thing we need to make more explicit are the effective auxiliary fields. To do so we 
start by deriving $W^{\rm eff}$ by substituting the solution (\ref{aphi}) into in $W$. Taking 
a derivative we then find that $W^{\rm eff}_i = W_i + \epsilon_i^\alpha W_\alpha$. But 
using the stationarity condition $W_{\alpha I} \bar W_{\bar I} = 0$ of the heavy scalars we 
see that $W_\alpha = \bar \epsilon_{\bar \imath}^{\bar \alpha} W_i$, so that
$W^{\rm eff}_i = (\delta_{i \bar \jmath} + \epsilon_i^\alpha \bar \epsilon_{\bar \jmath}^{\bar \alpha}) W_j
= g_{i \bar \jmath}^{\rm eff} W_j$. The auxiliary fields in the effective theory thus coincide with the 
light components of the auxiliary fields in the microscopic theory: 
$F^{i{\rm eff}} = - g^{{\rm eff}i \bar \jmath} \bar W_{\bar \jmath}^{\rm eff} = - \bar W_{\bar \imath} = F^i$.
Recalling (\ref{aF}) one also finds that  $g^{\rm eff}_{i \bar \jmath} F^{i{\rm eff}} \bar F^{\bar \jmath{\rm eff}} = F^I \bar F^{\bar I}$.
In summary, we get:
\be
F^{i{\rm eff}} = F^i \,,\;\; F^{i{\rm eff}} \bar F_i^{{\rm eff}} = F^I \bar F_I \,.
\ee
To proceed, we also need to compute more explicitly the mass-matrix blocks (\ref{f1}) and (\ref{f2})
entering in the expression (\ref{mibj}) for the effective mass matrix $m^{2\e}_{0 i \bar \jmath}$. 
In normal coordinates, these quantities depend on $|M^\epsilon|^2_{\alpha \bar \beta} =  M_{\alpha \gamma} 
\big(g^{\gamma \bar \delta} + \epsilon_i^\gamma \bar \epsilon_{\bar \imath}^{\bar \delta}
\big) \bar M_{\bar \delta \bar \beta}$, and at quadratic order in the auxiliary fields one finds that 
\bea
V_{\alpha \bar \beta} \b=\b |M^\epsilon|^2_{\alpha \bar \beta} - R_{\alpha \bar \beta K \bar L} F^K \bar F^{\bar L} \,,\;\;
V_{\alpha \beta} = - \lambda_{\alpha \beta K} F^K \,, \label{c1}\\
V_{\rm inv}^{\alpha \bar \beta} \b=\b |M^\epsilon|^{-2 \alpha \bar \beta} 
+ |M^\epsilon|^{-2 \alpha \bar \delta} |M^\epsilon|^{-2 \bar \beta \gamma} 
R_{\gamma \bar \delta K \bar L} F^K \bar F^{\bar L} \,. \label{c2}
\eea
We are now in position to evaluate the average sGoldstino mass in the effective theory by computing the four
correction terms in eq.~(\ref{mibj}). As explained after eqs.~(\ref{mibj}) and (\ref{mij}), these give rise to three types
of effects. But when looking along the sGoldstino direction, some simplifications occur, due to the fact that only
supersymmetry-breaking effects matter. The first type of effect cancels the corresponding leading part of 
$m_{0 i \bar \jmath}^2$. The second type of effect combines with the corresponding subleading term in 
$m_{0 i \bar \jmath}^2$ to reconstruct the average sGoldstino mass of the microscopic theory. The third 
type of effect gives instead a genuine correction. 
The precise evaluation of these effects can be simplified by noticing that at a stationary point $W_{I J} \bar W_{\bar J} = 0$,
which implies that at leading order in the auxiliary fields $V_{\alpha \bar \imath} W_i = - V_{\alpha \bar \beta} W_\beta$. 
After a straightforward computation one finds that $m_\varphi^{2{\rm eff}} = 
- \big(R_{I \bar J K \bar L} + \lambda_{\alpha I K} |M^\epsilon|^{-2 \alpha \bar \beta} 
\bar \lambda_{\bar \beta \bar J \bar L} \big)F^I \bar F^{\bar J} F^K \bar F^{\bar L}/F^M \bar F_M$.
Recalling then that $F^\alpha = \epsilon^\alpha_i F^{i {\rm eff}}$ and $F^I \bar F_I = F^{i {\rm eff}} \bar F_{i}^{\rm eff}$, 
one may finally rewrite the above result as 
\bea
m_\varphi^{2{\rm eff}} = \big(R^\epsilon - \lambda_\alpha^\epsilon |M^\epsilon|^{-2 \alpha \bar \beta} \bar \lambda_{\bar \beta}^\epsilon \big) 
F^{i{\rm eff}} \bar F_i^{\rm eff} \,, \label{m2}
\eea
with
\bea
\a\a R^\epsilon = - \frac {R^\epsilon_{i \bar \jmath k \bar l} F^{i{\rm eff}} \bar F^{\bar \jmath{\rm eff}} F^{k{\rm eff}} \bar F^{\bar l{\rm eff}}}
{(F^{m{\rm eff}} \bar F_m^{\rm eff})^2} \,, \\
\a\a \lambda_\alpha^\epsilon = \frac {\lambda^\epsilon_{\alpha ij} F^{i{\rm eff}} F^{j {\rm eff}}}{F^{k{\rm eff}} \bar F_k^{\rm eff}} \,.
\eea
The first term in the result (\ref{m2}) corresponds to $m^2_\varphi$, whereas the second term describes a negative 
level-repulsion effect controlled by the Yukawa couplings $\lambda_{\alpha i j}$ mixing one heavy and two light fields. 
As anticipated, the dependence on $\epsilon$ amounts to a transformation of all the tensorial quantities accounting 
for the need to disentangle light from heavy eigenmodes of the supersymmetric mass matrix, and can thus be dropped 
by setting $\epsilon$ to zero.

\subsection{Superfield approach}

The above results can also be derived by integrating out the heavy fields directly at the superfield level,
and then computing the sGoldstino mass in the resulting effective theory by applying eqs.~(\ref{meff1}) and 
(\ref{meff2}). To do this, one derives the effective K\"ahler potential and superpotential by solving the following 
approximate superfield equations of motion:
\be
\Phi^\alpha = \Phi^\alpha(\Phi^i) \;\;\text{solution of}\;\; W_\alpha (\Phi^i,\Phi^\alpha)= 0 \label{bphi}\,.
\ee
The bosonic components of this superfield equations of motion coincide, at leading order in the number 
of fermions and auxiliary fields, with the equations of motion (\ref{aphi})--(\ref{aF}) that we have used in the 
component approach. 

To proceed, we will need to compute the first and second derivatives of the heavy 
scalar fields with respect to the light scalar fields. These can be derived by differentiating eq.~(\ref{bphi}),
and one finds the following results:
\be
\frac {\partial \phi^\alpha}{\partial \phi^i}  = \epsilon^\alpha_i \,,\;\;
\frac {\partial^2 \phi^\alpha}{\partial \phi^i \partial \phi^j} = 
- M^{-1\alpha \beta} \lambda^\epsilon_{\beta i j} \,. \label{der}
\ee
The effective geometry can be derived by taking derivatives with respect to the light fields of the effective
K\"ahler potential $K^{\rm eff}$, where the heavy fields have been substituted by the solution (\ref{bphi}) 
in terms of light fields. We focus again on a given point in the light field space, around which we choose normal 
coordinates, but this point no longer needs to be a stationary point. Then, using the chain rule and 
eqs.~(\ref{der}), one easily computes
$K_{i \bar \jmath}^{\rm eff} = \delta_{i \bar \jmath} + \epsilon^\alpha_i \bar \epsilon^{\bar \alpha}_{\bar \jmath}$,
$K^\e_{\bar\imath j k} = - M^{-1 \alpha \beta} \bar \epsilon^{\bar \alpha}_{\bar \imath} \lambda^\epsilon_{\beta j k}$ 
and $K^\e_{i\bar\jmath k\bar l} = R^\epsilon_{i\bar\jmath k\bar l} 
+ \lambda^\epsilon_{\alpha i k} |M|^{-2 \alpha \bar \beta} \bar \lambda^\epsilon_{\bar \beta \bar \jmath \bar l}$.
This finally implies that the effective metric is given by $g_{i \bar \jmath}^{\rm eff} = g^\epsilon_{i \bar \jmath}$,
the effective Christoffel symbol by 
$\Gamma^\e_{\bar\imath j k} = - M^{-1 \alpha \beta} \bar \epsilon^{\bar \alpha}_{\bar \imath} \lambda^\epsilon_{\beta j k}$
and finally the effective Riemann tensor by the following expression:
\bea
\a\a R^\e_{i\bar\jmath k\bar l} = R^\epsilon_{i\bar\jmath k\bar l} 
+ \lambda^\epsilon_{\alpha i k} |M^\epsilon |^{-2 \alpha \bar \beta} \bar \lambda^\epsilon_{\bar \beta \bar \jmath \bar l} \,
\label{Riemann}
\eea
Plugging this expression into eqs.~(\ref{meff1}) and (\ref{meff2}), we then reproduce the form of the result 
(\ref{m2}).\footnote{Note that the results derived in this subsection are evaluated at values of the heavy scalar fields 
solving $W_\alpha = 0$, whereas the results of previous section were evaluated at values of the 
heavy scalar fields solving $V_\alpha = 0$. However it turns out that the difference between 
these two values is subleading in the counting of auxiliary fields and can therefore be discarded.}

\section{Models with chiral and vector multiplets}
\setcounter{equation}{0}

Let us now consider the case of $N=1$ theories with chiral multiplets $\Phi^I$
and vector multiplets $V^a$. The most general two-derivative Lagrangian 
is in this case specified by a real K\"ahler potential $K$, a holomorphic superpotential 
$W$, a holomorphic gauge kinetic function $f_{ab}$ and some holomorphic Killing vectors 
$X_a^I$:\footnote{We omit for simplicity the possibility of Fayet-Iliopoulos terms for 
Abelian factors.}
\be
\mathcal{L} = \int \! d^4 \theta\, K(\Phi,\bar \Phi,V)
+ \int \! d^2 \theta\, \Big[W(\Phi) + \s{\frac 14}\, f_{ab} (\Phi) \,W^{a \alpha} W_\alpha^b \Big] + {\rm h.c.} \,.
\ee
The gauge transformations of the chiral multiplets are defined by the Killing vectors $X_a^I$
whereas those of the vector superfields depend only on the structure constants $f_{ab}^{\;\;\;c}$
of the gauge group. Gauge invariance of the Lagrangian imposes that the variation of the non-holomorphic terms 
should be at most a K\"ahler transformation of the form $\Lambda^a f_a + \bar \Lambda^a \bar f_a$, 
where the $f_a$ are some holomorphic functions, whereas the holomorphic terms should be 
strictly invariant. This implies the following conditions:
\bea
\a\a X_a^I K_I - \s{\frac i2} K_a = f_a \,, \\[-0.5mm]
\a\a X_a^I W_I = 0 \,, \label{Winv} \\[0.5mm]
\a\a X_a^I f_{bcI} = - 2 f_{a\t{(}b}^{\;\;\;\;d} f_{c\t{)}d} \,.
\eea
These equations show that $-\frac 12 K_a$ can be identified with the real Killing potential
for the Killing vector $X_a^I$. They also imply that $K_{aI} = 2i \bar X_{aI}$ and 
$K_{ab} = 4 g_{I \bar J} X_{\t{(}a}^I \bar X_{b\t{)}}^{\bar J}$.
Finally, the equivariance condition on the Killing vectors guarantees 
that the Killing potentials can be chosen to transform in the adjoint representation, 
so that
\be
g_{I \bar J} X_{\t{[}a}^I \bar X_{b\t{]}}^{\bar J} = \s{\frac i4}\, f_{ab}^{\;\;\; c} K_c 
\label{equiv}\,.
\ee
In components and in the Wess-Zumino gauge, one finds $\mathcal{L} = T - V$ where:
\bea
T \b=\b - g_{I \bar J} \, D_\mu \phi^I D^\mu \bar \phi^{\bar J}
- \s{\frac 14} h_{ab} \, F^a_{\mu\nu} F^{b\mu\nu} + \s{\frac 14} \theta_{ab} \, F^a_{\mu\nu}\tilde{F}^{b\mu\nu} \nn \\
\b\;\b -\,i g_{I \bar J} \, \psi^I \big(D\!\!\!\!/\,\bar \psi^{\bar J} + 
\Gamma^{\bar J}_{\bar M \bar N} \, D\!\!\!\!/\, \bar \phi^{\bar M} \bar \psi^{\bar N} \big)
-\frac i2 h_{ab} \, \lambda^{a} D\!\!\!\!/\, \bar \lambda^b + {\rm h.c.} \nn \\
\b\;\b +\, \s{\frac 1{\!\sqrt{2}}} h_{abI} \, \lambda^a \sigma^{\mu \nu} \psi^I F^{b}_{\mu\nu} + {\rm h.c.} \,, \\ 
V \b=\b g^{I \bar J} \, W_I \bar W_{\bar J} + \s{\frac 18} \, h^{ab} K_a K_b \nn \\[-0.5mm]
\b\;\b + \s{\frac 12} \Big[\nabla_I W_J \, \psi^I \psi^J - g^{I \bar J} h_{abI} \bar W_{\bar J} \, \lambda^a \lambda^b
+ \s{\sqrt{8}} \big(g_{I \bar J} \bar X_a^{\bar J} + \s{\frac i4} h^{bc} h_{abI} K_c \big) \psi^I \lambda^a \Big] + {\rm h.c.} \nn \\
\b\;\b - \s{\frac 14} R_{I \bar J K \bar L} \, \psi^I \psi^K \bar \psi^{\bar J} \bar \psi^{\bar L}
+ \s{\frac 14} g^{I \bar J} h_{abI} h_{cd \bar J} \, \lambda^a \lambda^b \bar \lambda^c \bar \lambda^d 
+ \s{\frac 12} h^{cd} h_{acI} h_{bd \bar J} \, \psi^I \lambda^a \bar \psi^{\bar J} \bar \lambda^b \nn \\
\b\;\b - \s{\frac 14} \Big[\nabla_I h_{abJ} \, \psi^I \psi^J \lambda^a \lambda^b + 
h^{cd} h_{acI} h_{bdJ} \psi^I \lambda^a \psi^J \lambda^b \Big] + {\rm h.c.} \,.
\eea
In these expressions $D_\mu$ is the covariant derivative acting as 
$D_\mu \phi^I = \partial_\mu \phi^I + A^a_\mu X_a^I$, $D_\mu \psi^I = \partial_\mu \psi^I + A^a_\mu \partial_J X_a^I \,\psi^J$
and $D_\mu \lambda^a = \partial_\mu \lambda^a + f_{bc}^{\;\;\;a} A_\mu^b \lambda^c$, whereas $F_{\mu\nu}^a$ is  
the field-strength $F^a_{\mu \nu} = \partial_\mu A^a_\nu - \partial_\nu A_\mu^a + f_{bc}^{\;\;\;a} A_\mu^b A_\nu^c$ and 
$h_{ab}$ and $\theta_{ab}$ denote the real and imaginary parts of $f_{ab}$. 

A vacuum is defined by constant values of the scalars $\phi^I$ and 
vanishing values of the fermions $\psi^I,\lambda^a$ and the vectors $A_\mu^a$, 
such that $V$ is stationary. Supersymmetry is  spontaneously broken whenever 
some of the auxiliary fields $F^I,D^a$ have non-vanishing values. The form of 
these auxiliary fields is given by
\bea
\a\a F^I = - \bar W^I \,,\\
\a\a D^a = - \s{\frac 12} h^{ab} K_b \,.
\eea
The stationarity condition implies that 
\be
\nabla_I W_J \,F^J  + \s{\frac 12} h_{abI} D^a D^b + i \bar X_{a I} D^a = 0 \,.
\ee
Moreover, by contracting this relation with the Killing vectors $X_a^I$ and taking the imaginary 
part, and using (\ref{Winv}) and its derivative as well as (\ref{equiv}), one finds the following 
relation between the values of $F^I$ and $D^a$:
\be
i \nabla_I X_{a \bar J} \, F^I \bar F^{\bar J} - g_{I \bar J} X_{\t{(}a}^I \bar X_{b\t{)}}^{\bar J}\, D^b 
+ \s{\frac 12} f_{ab}^{\;\;\;d} \theta_{dc} \,D^b D^c = 0 \,. \label{relDFF}
\ee
The masses of the scalar, fermion and vector fields describing fluctuations around 
the vacuum are found to be given by
\bea
m^2_{0 I \bar J} \b=\b \nabla_I W_K \nabla_{\bar J} \bar W^K - R_{I \bar J K \bar L} \, F^K \bar F^{\bar L} 
+ h^{ab} \bar X_{a I} X_{b \bar J} + h^{ab} h_{acI} h_{bd \bar J} \, D^b D^c \nn \\[1mm]
\b\;\b +\, \big(q_{a I \bar J} - i h^{bc} h_{abI} X_{c \bar J} + i h^{bc} h_{ab \bar J} \bar X_{c I} \big) D^a \,, 
\label{m0vec}\\[0mm]
m^2_{0IJ} \b=\b - \nabla_I \nabla_J W_K \, F^K - h^{ab} \bar X_{a I} \bar X_{b J} 
- \s{\frac 12} \big(\nabla_I h_{abJ} - 2 h^{cd} h_{acI} h_{bdJ} \big) \, D^a D^b \nn \\[0.5mm]
\b\;\b +\,2 i\, h^{bc} h_{ab\t{(}I} \bar X_{cJ\t{)}} D^a \,,
\eea
then
\bea
\a\a m_{1/2 IJ} = \nabla_I W_J \,, \\[1.5mm]
\a\a m_{1/2 ab} = h_{abI} \, F^I \,, \\[0mm]
\a\a m_{1/2 Ia} = \s{\sqrt{2}}\, \bar X_{aI} - \s{\frac i{\!\sqrt{2}}} h_{abI} \, D^b \,,
\eea
and finally
\bea
\a\a m^2_{1 ab} = 2\, g_{I \bar J} X_{\t{(}a}^I \bar X_{b\t{)}}^{\bar J} \,.
\eea
We see that the supersymmetric parts of the mass matrices are given by $W_{IJ}$ for the chiral multiplets 
and by $2 g_{I \bar J} X_{\t{(}a}^I \bar X_{b\t{)}}^{\bar J}$ for the vector multiplets.

The directions $F^I$ and $D^a$ in field space are special. For fermions they define the Goldstino 
$\eta = \bar F_I \psi^I + \frac i{\sqrt{2}} D_a \lambda^a$, which is massless and represents 
the Goldstone mode of broken supersymmetry: $m_\eta = 0$.
For scalars they define instead the projected sGoldstino $\varphi = \bar F_I \phi^I$, which 
describes two real scalar fields with masses that are entirely controlled by supersymmetry breaking 
effects. One may then consider the average of these two masses, which is as before given by
\be
m_\varphi^2 = \frac 
{m_{0I \bar J}^{2} F^{I} \bar F^{\bar J}}
{F^K \bar F_K} \,.
\label{mmicvec}
\ee
A straightforward computation shows that the result is in this case given by \cite{GRS3,JS}
\be
m_{\varphi}^2 = R \,F^I \bar F_I + S\, D^a D_a + \s{\frac 14}\, T\, \frac {(D^a D_a)^2}{F^I \bar F_I} 
+ M^2 \frac {D^a D_a}{F^I \bar F_I} \,,
\ee
where
\bea
\a\a R = - \frac {R_{I \bar J K \bar L} \,F^I \bar F^{\bar J} F^K \bar F^{\bar L}}{(F^M \bar F_M)^2} \,, \label{RFD} \\
\a\a S = \frac {h_{acI} h^{cd} h_{db \bar J} \,F^I \bar F^{\bar J} D^a D^b }{(F^K \bar F_K)(D^c D_c)} \,, \label{SFD} \\
\a\a T = \frac {h_{abI} h_{cb}^{\;\;\;I} D^a D^b D^c D^d}{(D^e D_e)^2} \,, \label{TFD} \\
\a\a M^2 = \frac {2 X_a^I \bar X_{bI}\, D^a D^b}{D^c D_c} \,. \label{MFD}
\eea

The directions $X_a^I$ in field space are also special. In the supersymmetric limit 
they define the chiral multiplets $\Phi^a$ that are eaten up by 
the massless vector multiplets to produce massive vector multiplets. When supersymmetry
is broken, things get a bit more complicated but the fermionic and bosonic components
of $\Phi^a$ still get a mass comparable to that of the components of $V^a$.

From the above result it follows that a necessary condition for not having a tachyonic 
mode is that the holomorphic sectional curvature $R$ be larger than some negative 
value depending on the gauge sector data \cite{GRS3}. In this case it is less clear 
whether this necessary condition becomes also sufficient if for a given $K$ one 
allows $W$ to be adjusted. Indeed, 
gauge-invariance forbids any tuning of $W_I$, $W_{IJ}$ and $W_{IJK}$ along 
the directions $X_a^I$. The corresponding modes thus represent a priori a left-over
danger of instability \cite{JS}. This danger does however disappear in the limit 
we are considering here where the vector masses are much larger than the 
supersymmetry breaking scale, since these modes then become very heavy.

\subsection{Integrating out heavy vector multiplets}

Let us now suppose that all the vector multiplets have a large supersymmetric
mass, much larger than the splittings induced by supersymmetry breaking.
We may then integrate out in a supersymmetric way the modes associated with 
the heavy vector multiplets, paying attention to the fact that in order to become 
massive they absorb the modes of some chiral multiplets. 
The relevant scales in this case are the supersymmetric mass 
matrix $2 g_{I \bar J} X_{\t{(}a}^I \bar X_{b\t{)}}^{\bar J} = \frac 12 K_{ab}$ 
of the heavy vector multiplets and the quantity $i X_{aI} = \frac 12 K_{a I}$ 
controlling the supersymmetric mixing between vector multiplets and chiral 
multiplets:
\be
M^2_{ab} = \frac 12 K_{ab} \,,\;\; \nu_{aI} = \frac 12 K_{aI} \,.
\ee
The couplings that are expected to be relevant are instead 
given by the cubic couplings in $K$, namely the generalized charges
\be
q_{a I \bar J} = - \frac 12 K_{a I \bar J} \,,\;\; q_{a b I} = - \frac 12 K_{a b I} \,,\;\; q_{abc} = - \frac 12 K_{abc} \,.
\ee

At leading order in the expansion in number of derivatives, fermions and auxiliary fields,
the low-energy effective theory for the light chiral multiplets can again be obtained 
in two different but equivalent ways. One may proceed in components and integrate 
out the heavy modes associated to the vector multiplets and the chiral multiplets that they 
absorb, by requiring stationarity of $V$ with respect to them. 
One may however also proceed in superfields and integrate out the heavy vector superfields 
by requiring stationarity of $K$ with respect to them. For convenience, 
we shall as before assume without loss of generality normal coordinates in the microscopic 
theory around the point under consideration. 

In analogy with what happens in the case of only chiral multiplets, we expect that the corrections
due to the supersymmetric mixing between heavy and light multiplets should be encoded in 
following parameter of dimension one:
\be
\delta^a_I = - M^{-2ab} \nu_{bI} \,.
\ee
In this case, such a parameter cannot be set to zero by a simple holomorphic field redefinition, 
because it corresponds to the non-holomorphic mixing between the heavy gauge fields and 
the corresponding real would-be Goldstone modes. However, it can be set to zero by making 
a suitable choice of gauge. With any difference choice of gauge, $\delta^a_I$ would be 
non-zero and the terms depending on it in the effective theory would take into account the 
mixing between light and heavy fields. By doing the computation in such a gauge one would 
presumably end up getting deformed versions of all the tensorial quantities for light fields, 
involving additional contributions where heavy indices are converted to light indices by $\delta^a_I$. 
We shall however refrain from keeping a general $\delta^a_I \neq 0$ and set $\delta^a_I = 0$ 
from the beginning by choosing the unitary gauge.

To perform the splitting between light and heavy fields and the gauge fixing more precisely,
we may start by splitting the chiral multiplets $\Phi^I$ into those that are orthogonal and 
those that are parallel to the Killing vectors $X_a^I$ evaluated at the point under consideration. 
This decomposition can be done more explicitly with the help of the parallel projector 
$P^I_{\;\;J} = 2 X_a^I M^{-2 ab} \bar X_{b J}$. We shall denote these two sets of fields 
respectively with $\Phi^i$ and $\Phi^a$. The orthogonal components $\Phi^i$ define 
the light chiral multiplets of the low-energy effective theory. The parallel components 
$\Phi^a$ are instead either heavy or eliminable through the gauge fixing. 

In the following, we shall follow the same logic as in the previous section and first 
compute within the component approach the average sGoldstino mass in the 
low-energy effective theory, defined at a stationary point as
\be
m_\varphi^{2{\rm eff}} = \frac 
{m_{0i \bar \jmath}^{2{\rm eff}} F^{i{\rm eff}} \bar F^{\bar \jmath{\rm eff}}}
{F^{k{\rm eff}} \bar F_k^{\rm eff}} \,.
\label{meffvec}
\ee
We will then reproduce the same result within the superfield approach by first computing 
the Riemann tensor of the effective theory at a generic point and then plugging it in the 
expression for the sGoldstino mass at a stationary point within the effective theory, which 
is given by
\be
m_\varphi^{2{\rm eff}} = R^{\rm eff} F^{i{\rm eff}} \bar F_i^{\rm eff} \,, \label{meff1vec}
\ee
in terms of an effective sectional curvature
\be
R^{\rm eff} = - \frac {R_{i \bar \jmath k \bar l}^{\rm eff} F^{i{\rm eff}} \bar F^{\bar \jmath{\rm eff}} F^{k{\rm eff}} \bar F^{\bar l{\rm eff}}}
{(F^{m{\rm eff}} \bar F_m^{\rm eff})^2} \,. \label{meff2vec}
\ee

\subsection{Component approach}

Let us first consider the component approach, where it is convenient to choose the 
Wess-Zumino gauge for the extra gauge symmetries implied by supersymmetry. 
For simplicity we shall as before focus on bosonic fields and discard fermions
since we are interested in scalar masses. The relevant bosonic heavy modes coming 
from $V^a$ and $\Phi^a$ are the following. In the vector multiplets $V^a$, the gauge 
fields $A_\mu^a$ contain heavy physical modes and should of course be considered. 
In the chiral multiplets $\Phi^a$, on the other hand, the modes $\sigma^a = {\rm Re} (\phi^a)$ 
correspond to the would-be Goldstone modes and can be eliminated by choosing the 
unitary gauge for the standard gauge symmetries, where the corresponding degrees of 
freedom are the longitudinal polarizations of the gauge bosons, whereas the modes 
$\rho^a = {\rm Im}(\phi^a)$ are physical and easily seen to have a mass comparable to 
that of the vector fields, so that they must be considered. 
At leading order in the low-energy expansion, the heavy bosonic fields $A_\mu^a$ and $\rho^a$ 
can then be integrated out by using the following approximate equations of motion:
\bea
\rho^a \b=\b \rho^a(\phi^i,\bar \phi^{\bar \imath}) \;\;\text{solution of}\;\; V_a (\phi^i, \bar \phi^{\bar \imath}, \rho^a) = 0 \,, \label{eqrho} \\
A_\mu^a \b=\b 0 \,. \label{eqA}
\eea
Concerning the auxiliary fields, notice that those coming from the parallel chiral 
multiplets automatically vanish, as a consequence of the gauge invariance of the 
superpotential (\ref{Winv}), whereas those of the vector multiplets are given by 
eq.~(\ref{relDFF}), which corresponds to the equation of motion of $\rho^a$ and 
reduces approximately to $q_{a I \bar J} F^I \bar F^{\bar J} - \frac 12 M^2_{ab} D^b = 0$. 
At leading order in the low-energy expansion one then finds:
\bea
\a\a F^a = 0 \label{eqF} \,, \\
\a\a D^a = 2\, M^{-2ab} q_{b i \bar \jmath} F^i \bar F^{\bar \jmath} \label{eqD}\,.
\eea
The effective theory for the light fields is finally obtained by substituting these 
expressions into the Lagrangian.

To derive the effective theory, one needs in principle to compute the derivatives 
of $\rho^a$ with respect to $\phi^i$. This can be deduced by taking a derivative of 
the stationarity condition for $\rho^a$ with respect to $\phi^i$. One then finds
a result that is inversely proportional to the mass matrix of $\rho^a$, which 
is approximately equal to that of the vectors, and directly proportional to the mass 
mixing between $\rho^a$ and $\phi^i$. This mixing can be computed explicitly
and after using the relation (\ref{Winv}) ensuring the gauge invariance of $W$,
as well as its first and second derivatives, one verifies that it contains only terms 
that are quadratic in the auxiliary fields 
or linear in the auxiliary fields but further suppressed by the ratio between light 
chiral masses and heavy vector mass, which must all be neglected. As a result,
one finds:
\be
\frac {\partial \rho^a}{\partial \phi^i} = 0 \,. \label{dr}
\ee

The effective K\"ahler metric of the light fields is not affected. Indeed, 
neither $A_\mu^a$ nor $\rho^a$ give any effect in the kinetic terms, as a 
consequence of eqs.~(\ref{eqA}) and (\ref{dr}). One thus simply finds:
\be
g^{\rm eff}_{i \bar \jmath}= g_{i \bar \jmath} \,. \\
\ee

The effective scalar mass matrices can be computed by taking into account both the direct
effect of the heavy modes on the microscopic mass evaluated in the light scalar directions 
$\phi^i$ and the indirect level-repulsion effect coming from the mass mixing with the heavy 
scalar directions $\rho^a$. It turns however out that the level-repulsion effect is negligible,
for essentially the same reasons as those leading to eq.~(\ref{dr}). We thus finally get: 
\bea
m^{2\e}_{0 i \bar \jmath} \b=\b m^2_{0 i \bar \jmath} \,, \\
m^{2\e}_{0 i j} \b=\b m^2_{0 i j} \,.
\eea
There is nevertheless a direct effect in the Hermitian block $m_{0 i \bar \jmath}^{2\e}$, 
which consists of two significant contributions in $m^2_{0 i \bar \jmath}$ coming from the couplings 
to heavy fields. The first contribution comes from plugging back the small but non-vanishing 
value of $D^a$ into the last term of (\ref{m0vec}). It is easily evaluated by 
using eq.~(\ref{eqD}), and one finds 
$q_{ai \bar \jmath} D^a = 2\, q_{a i \bar \jmath} M^{-2ab} q_{b k \bar l} F^k \bar F^{\bar l}$.
The second contribution arises instead from the part of the first term in (\ref{m0vec}) that corresponds 
to values for the summed index $K$ that run over the parallel chiral modes that are integrated out. 
It can be evaluated by using the projected metric $P^{I \bar J} = 2 X_a^I M^{-2 ab} \bar X_b^{\bar J}$, 
and reads $\nabla_i W_a \nabla_{\bar \jmath} \bar W^a = 
\nabla_i W_K P ^{K \bar L} \nabla_{\bar \jmath} \bar W_{\bar L} 
= 2 X_a^K \nabla_i W_K M^{-2ab} X_b^{\bar L} \nabla_{\bar \jmath} \bar W_{\bar L}$. 
But taking a derivative of eq.~(\ref{Winv}) one deduces that 
$X_a^K \nabla_i W_K = - i q_{a i \bar K} \bar F^{\bar K} =  - i q_{a i \bar k} \bar F^{\bar k}$,
and finally $\nabla_i W_a \nabla_{\bar \jmath} \bar W^a 
= 2\, q_{a i \bar l} M^{-2ab} q_{b k \bar \jmath} F^k \bar F^{\bar l}$.
These two contributions represent a direct correction to all the masses, which may be either positive 
or negative depending on the value of charges along the direction that is considered.

Let us now evaluate more precisely the average sGoldstino mass defined by eq.~(\ref{meffvec}) 
at a stationary point of the effective theory and compare it to its analogue defined by eqs.~(\ref{meff1vec})
and (\ref{meff2vec}) in the microscopic theory. Along the supersymmetry breaking direction 
$F^{i \e} = F^i$ the two direct corrections discussed above give identical contributions that sum up 
and one easily finds:
\be
m_{\varphi}^{2\e} = \big(R + 4\, q_a M^{-2ab} q_b \big) \,F^{i \e} \bar F_i^\e \,, \label{m2vec}
\ee
where
\bea
\a\a R = - \frac {R_{i \bar \jmath k \bar l} \,F^{i \e} \bar F^{\bar \jmath \e} F^{k \e} \bar F^{\bar l \e}}
{(F^{m \e} \bar F_m^\e)^2} \,, \\
\a\a q_a = \frac {q_{a i \bar \jmath}\, F^{i\e} \bar F^{\bar \jmath \e}}{F^{k \e} \bar F_k^\e} \,.
\eea
The first term in the result (\ref{m2vec}) corresponds to $m^2_\varphi$, whereas the second term 
describes a positive direct effect controlled by the charges $q_{a i \bar \jmath}$ mixing one 
heavy and two light fields. The absence of any indirect level-repulsion effect is due to the absence 
of genuine heavy chiral multiplets mixing to the light chiral multiplets. 

\subsection{Superfield approach}

It is straightforward to show that the above results can also be obtained by integrating out 
the heavy vector multiplets at the level of superfields.
The only complication is that one should switch from the unitary plus Wess-Zumino gauge
used in the component formulation, which fix respectively the standard and the extra gauge 
symmetries, to a supersymmetric unitary gauge to be used in the superfield formulation, 
which fixes at once all the multiplet of gauge symmetries. More precisely, we shall gauge 
fix all the parallel chiral multiplets $\Phi^a$ to constant values coinciding with their values 
at the stationary point. The superfields $V^a$ become however general vector superfields in 
this gauge, and compared to the Wess-Zumino gauge that was chosen in the component 
approach, the modes that were described by the real scalar fields $\rho^a$ in the $\Phi^a$ 
have now been transfered to the real scalar fields $c^a$ in the general $V^a$. In this 
supersymmetric gauge, all the heavy degrees of freedom are thus contained in $V^a$, 
and can be integrated out by using the following approximate superfield equations of motion:
\be
V^a = V^a(\Phi^i,\bar \Phi^{\bar \imath}) \;\;\text{solution of}\;\; K_a(\Phi^i,\bar \Phi^{\bar \imath},V^a) = 0 \,. \label{cV}
\ee
The bosonic components of this superfield equations of motion map to the equations of motion (\ref{eqrho})--(\ref{eqD}) 
that we have used in the component approach, modulo the different gauge choice. 

To proceed, we will need to compute the first and second derivatives of the lowest component of the heavy 
vector superfields with respect to the light scalar superfields. These can be derived by differentiating 
eq.~(\ref{cV}), and at the point under consideration where $K_{ai} = 0$ one finds the following results:
\be
\frac {\partial c^a}{\partial \phi^i} = 0 \,,\;\; 
\frac {\partial^2 c^a}{\partial \phi^i \partial \bar \phi^{\bar \jmath}} = M^{-2 ab} q_{b i \bar \jmath} \,. 
\label{dervec}
\ee
The effective geometry can be derived by taking derivatives with respect to the light fields of the effective
K\"ahler potential $K^{\rm eff}$, where the heavy fields have been substituted by the solution (\ref{cV}) 
in terms of light fields. We focus again on a given point in the light field space and use normal coordinates.
Then, using the chain rule and eq.~(\ref{dervec}), and noticing that $K_{aij} = 0$ and 
$K_{a i \bar \jmath} = -2 q_{a i \bar \jmath}$, one easily computes 
$K_{i \bar \jmath}^{\rm eff} = \delta_{i \bar \jmath}$, $K^\e_{\bar\imath j k} = 0$ and 
$K^\e_{i\bar\jmath k\bar l} = K_{i\bar\jmath k\bar l} 
-2\, q_{a i \bar \jmath} M^{-2 ab} q_{b k \bar l} -2\, q_{a i \bar l} M^{-2 ab} q_{b k \bar \jmath}$.
This finally implies that the effective metric is given by $g_{i \bar \jmath}^{\rm eff} = g_{i \bar \jmath}$,
the effective Christoffel symbol by $\Gamma^\e_{\bar\imath j k} = 0$ and the effective Riemann tensor
by the following expression:
\bea
\a\a R^\e_{i\bar\jmath k\bar l} = R_{i\bar\jmath k\bar l} 
- 2\, q_{a i \bar \jmath} M^{-2 ab} q_{b k \bar l} 
- 2\, q_{a i \bar l} M^{-2 ab} q_{b k \bar \jmath} \,. \label{Riemannvec}
\eea
Plugging this expression into eqs.~(\ref{meff1vec}) and (\ref{meff2vec}), we then reproduce the form 
of the result (\ref{m2vec}). 

\section{Conclusion}
\setcounter{equation}{0}

Summarizing, we have shown that integrating out heavy chiral multiplets $\Phi^\alpha$ and vector 
multiplets $V^a$ with large and approximately supersymmetric mass matrices $M^{2 \alpha \bar \beta}$ 
and $M^{2 ab}$ induces corrections to the square masses of light scalars $\phi^i$ that are due 
respectively to an indirect level-repulsion effect and a direct coupling effect. The 
crucial dimensionless couplings that are involved in these effects are respectively the Yukawa
couplings $\lambda_{\alpha i j} = W_{\alpha i j}$ and the generalized gauge charges 
$q_{a i \bar \jmath} = - \frac 12 K_{a i \bar \jmath}$, which corresponds to cubic couplings mixing one heavy 
and two light multiplets respectively in $W$ and $K$. In particular, by looking along the chiral 
projection of the supersymmetry breaking direction, which is defined by the chiral auxiliary 
fields $F^i$, we showed that the averaged sGoldstino mass in the effective theory takes the form:
\be
m_{\varphi}^{2\e} = \big(R - \lambda_\alpha |M|^{-2 \alpha \bar \beta} \bar \lambda_{\bar \beta} 
+ 4\, q_a M^{-2 ab} q_b \big) M_{\rm S}^4 \,. \label{res}
\ee
The first term is what one would find by just restricting to the light fields. It is controlled by 
the sectional curvature $R$ along the $F$-direction, and can have any sign. The second 
term is the correction induced by heavy chiral multiplets. It is controlled by the Yukawa
couplings $\lambda_\alpha$ along the $F$-direction and is always negative. The third term 
is the correction induced by heavy vector multiplets. It is controlled by the gauge charges 
$q_a$ along the $F$-direction and is always positive. Finally $M_S$ is the scale of 
supersymmetry breaking, which in our situation is set by the $F$ auxiliary fields since 
the $D$ auxiliary fields are suppressed.

The result (\ref{res}) has been derived in rigid supersymmetry, in the limit where the supersymmetry
breaking scale is much lower than the mass scale $M$ of the heavy modes that are integrated out. 
Its generalization to gravity can however be derived in a straightforward way by using the results
of \cite{BGRS}, where it was shown that whenever the gravitino mass $m_{3/2}$ is also much smaller 
than the heavy mass scale $M$, one may first integrate out the fields in the rigid limit and then switch 
on the coupling to gravity. More precisely, the only modification induced by gravity in (\ref{res}) is 
the addition of the correction $2 m_{3/2}^2$, which reconstructs the supergravity result of 
\cite{GRS1} for the sGoldstino mass in the theory truncated to light modes:
\be
\Delta m_\varphi^{2{\rm eff}} = 2\, m_{3/2}^2 \,. \label{dres}
\ee

The origin of the difference in sign in the corrections induced by heavy chiral and vector multiplets
is transparent in the component approach, where the first is due to an indirect level-repulsion effect 
whereas the second is due to a direct coupling effect. In the superfield approach, the two 
computations look instead very symmetric and the difference in sign is at first sight surprising. A closer inspection 
shows however that there too it can be understood quite robustly. For this we observe that for heavy chiral 
multiplets the stationarity condition $W_\alpha = 0$, the auxiliary fields $\bar F_\alpha = - W_\alpha$ and 
the relevant cubic couplings $\lambda_{\alpha i j} = W_{\alpha i j}$ are all controlled by the superpotential $W$,
whereas for heavy vector multiplets the stationarity condition $K_a = 0$, the auxiliary fields $D_a = - \frac 12 K_a$ 
and the relevant cubic couplings $q_{a i \bar \jmath} = - \frac 12 K_{a i \bar \jmath}$ are all controlled by the K\"ahler 
potential $K$. There is then a perfect symmetry between the two dynamics, which exchanges the roles of $K$ 
and $W$. When one looks at the effects of these heavy dynamics onto the supersymmetry-breaking part 
of the masses of light scalar fields, this symmetry is however broken, because supersymmetry-breaking 
contributions to scalar masses arise only from $K$ and not from $W$. This is what causes 
the difference in sign between the two effects.\footnote{A similar phenomenon has also been 
encountered in different context in \cite{DFsoft}.}

The result that we have obtained may have interesting applications in the context of string models, 
where the situation in which some of the multiplets are stabilized in a supersymmetric way at a
high energy scale naturally occurs and the question of their effect on the dynamics of 
the light multiplets, which are supposed to break supersymmetry, acquires a crucial 
importance.
In such a situation one has in principle to honestly integrate out the heavy fields to properly
describe the dynamics of the light fields. But it is in general cumbersome to do so, and this
raises the question of whether or when one may get a qualitatively reliable indication on the 
light field dynamics by just freezing the heavy fields and truncating the theory. Some particular
situations where one can safely do this truncation and get the right effective theory have 
been identified in \cite{AHS,GS1,GS2}. Here we have shown more specifically and more 
systematically what kind of dangers may arise from the heavy fields concerning the masses 
of the light fields, which are the crucial issue for metastability of the vacuum.

A concrete example is that of string models where large classical effects related to background 
fluxes stabilize some moduli in a supersymmetric way with a large mass and small quantum 
effects related to gauge interactions stabilize some other moduli in a non-supersymmetric 
way with a small mass \cite{GKP,KKLT}. The dynamics of these heavy and light modes, schematically
denoted by $H$ and $L$, is then described by $K = K_L (L,\bar L) + K_H (H,\bar H) + K_Q(L, \bar L, H, \bar H)$ 
and $W = 0 + W_H(H) + W_Q(L,H)$. For gauge interactions with a field-dependent gauge kinetic 
function $f \propto L$, the quantum effects have the following structure. The correction $K_Q$ 
consists of both perturbative and non-perturbative effects suppressed by inverse powers and exponentials 
of $L + \bar L$, and can usually be neglected, since it represents a small correction to the kinetic 
terms of $L$. The correction $W_Q$ consists instead only of non-perturbative effects suppressed
by exponentials of $L$, and must be kept, since it represents the  dominant source of potential for 
$L$.\footnote{See \cite{GC1,GC2,GC3} for a more detailed discussion of these
effects for gaugino condensation.} In this situation, freezing the heavy moduli $H$ to constant 
values is a priori not justified \cite{CF,dA2,AHK}, but turns out a posteriori to give a sensible 
approximation to the effective theory for the light moduli $L$ thanks to the smallness of the 
quantum corrections mixing $L$ and $H$ \cite{GS1}. Applying our general result, we may now 
establish more quantitatively the importance of the corrections induced by integrating out the 
heavy modes on the light masses, and in particular the sGoldstino mass $m$. The relevant 
Yukawa coupling $\lambda$ between one $H$ and to $L$ fields will involve the same 
exponential suppression factor as $W_Q$. The dangerous indirect level-repulsion effect on 
$m^2$ will then be suppressed by the square of this exponential factor. On the other hand, 
the direct effect induced on $m^2$ from the mixing $K_Q$ involves both power and exponentially 
suppressed corrections. Given then that in these models there is a unique ultraviolet mass scale 
around $M_{\rm Pl}$, the indirect effect is a priori smaller than the direct effect, and in all 
the situations where the direct effect is neglected also the indirect effect must be discarded. 
There is thus no problem in the limit of small quantum effects.

One may finally wonder whether integrating out heavy chiral and vector multiplets has similar 
effects on soft masses in scenarios where both the visible and the hidden sectors couple to them.
In fact, these effects are easily computed, since they are also encoded in the effective Riemann 
tensor, but with two visible-sector and two hidden-sector indices: 
$m_{u \bar v}^{2{\rm eff}} = - R_{u \bar v i \bar \jmath}^{\rm eff} F^{i \e} \bar F^{\bar \jmath \e}$. 
Applying the results (\ref{Riemann}) and (\ref{Riemannvec}) one would then find 
\be
m^{2\e}_{u \bar v} = - \big(R_{u \bar v i \bar \jmath} 
+ \lambda_{\alpha u i} |M|^{-2 \alpha \beta} \bar \lambda_{\bar \beta \bar v \bar \jmath} 
- 2\, q_{a u \bar v} M^{-2 ab} q_{b i \bar \jmath} 
- 2\, q_{a u \bar \jmath} M^{-2 ab} q_{b i \bar v}  \big) F^i \bar F^{\bar \jmath} \,. \label{soft}
\ee
The first term is the usual expression for the soft masses, the second term represents 
the correction induced by heavy chiral multiplets, and the third and fourth terms describe 
the corrections induced by heavy vector multiplets. The various couplings controlling these 
effects are however not always allowed by the Standard Model gauge symmetry $G_{\rm SM}$. 
If the heavy states are neutral, only $q_{a u \bar v}$ and $q_{a i \bar \jmath}$ can be non-zero. The only 
effect then comes from the third term, with an arbitrary sign. This is the standard effect 
induced by a neutral heavy vector multiplet.\footnote{See for example \cite{ADM}.}
If on the other hand the heavy 
states are charged, only $\lambda_{\alpha u i}$ and $q_{a u \bar \jmath}$ can be non-zero.
The only effects then come from the second and the fourth terms, which are respectively 
negative and positive. However a charged chiral multiplet cannot have a supersymmetric 
mass term, because $G_{\rm SM}$ does not allow holomorphic invariants, whereas a charged 
vector multiplet can, since non-holomorphic invariants exist; so actually only the fourth term is relevant.
This is a less-standard but already-known effect that can be induced by charged vector multiplets.\footnote{This 
kind of effect is relevant in Grand Unified Theories, where charged massive vector fields occur after 
the gauge symmetry is broken down from $G_{\rm GUT}$ to $G_{\rm SM}$, and induces important corrections 
to soft masses. This phenomenon and its phenomenological implications were studied in detail in \cite{PD,R}.} 
In addition to these effects, there is as usual a separate gravitational effect, which for 
generic cosmological constant $V = M_{\rm S}^4 - 3 m_{3/2}^2 M_{\rm Pl}^2$ is given by:
\be
\Delta m^{2\e}_{u \bar v} = g_{u \bar v} \, \big(m_{3/2}^2 + V M_{\rm Pl}^{-2}\big)\,. \label{dsoft}
\ee
We clearly see that eqs.~(\ref{soft}) and (\ref{dsoft}) for the soft scalar masses correspond to 
eqs.~(\ref{res}) and (\ref{dres}) for the average sGoldstino mass.

\vskip 20pt

\noindent
{\bf \Large Acknowledgements}

\vskip 10pt

\noindent
This work was partly supported by the Swiss National Science Foundation. 
It is a pleasure to thank M.~G\'omez-Reino and R.~Rattazzi for useful comments
and discussions.

\end{document}